\documentclass[a4paper]{article}
\usepackage{amsmath,amsfonts,amssymb,amsthm}
\newcommand{\R}{{\mathbb R}}

\newcommand{\p}{{\bf p}}
\newcommand{\ggamma}{\vec{\gamma}}

\newcommand{\y}{{\bf y}}

\newcommand{\x}{{\bf x}}
\newcommand{\bc}{{\bf c}}
\newcommand{\A}{{\bf A}}
\newcommand{\ba}{{\bf a}}

\newcommand{\n}{{\bf n}}
\newcommand{\bk}{{\bf k}}
\numberwithin{equation}{section}
\theoremstyle{plain}
\newtheorem{theorem}{Theorem}[section]

\begin{document}

\noindent 
\begin{center}
\textbf{\large The Faraday effect revisited: General theory}
\end{center}

\begin{center}
14th of November, 2005
\end{center}

\vspace{0.5cm}

\noindent 

\begin{center}
\textbf{ 
Horia D. Cornean\footnote{Dept. Math., 
    Aalborg
    University, Fredrik Bajers Vej 7G, 9220 Aalborg, Denmark; e-mail:
    cornean@math.aau.dk},
Gheorghe Nenciu\footnote{Dept. Theor. Phys.,
University of Bucharest, P. O. Box MG11, RO-76900 Bu\-cha\-rest, Romania; }
\footnote{
Inst. of Math. ``Simion Stoilow'' of
the Romanian Academy, P. O. Box 1-764, RO-014700 Bu\-cha\-rest, Romania;
e-mail: Gheorghe.Nenciu@imar.ro},  
Thomas G. Pedersen \footnote{Dept. Phys. and Nanotech., Pontoppidanstr{\ae}de
  103, 9220 Aalborg, Denmark; 
e-mail: tgp@physics.aau.dk}}
     
\end{center}

\vspace{0.5cm}

\noindent

\begin{abstract}
 This paper is the first in a series revisiting 
   the Faraday effect, or more generally, the theory of electronic quantum
   transport/optical response in  bulk media in the presence of a constant
 magnetic field. The independent electron approximation is assumed. At
   zero temperature and zero frequency, if the Fermi energy lies in a
   spectral gap, we rigorously prove the Widom-Streda formula. 
 For free electrons, the
 transverse conductivity can be explicitly computed and coincides
 with the classical result. In the general case, using magnetic perturbation
theory, the conductivity tensor is expanded in powers of the strength
of the magnetic field $B$. Then the linear term in $B$ of this expansion is
written down  in terms of the zero magnetic field Green function and
   the zero field current operator. 
In the periodic case, the linear term in $B$ of the conductivity tensor is
expressed in terms of zero magnetic field Bloch functions and
energies. No derivatives with respect to the quasi-momentum appear and
thereby all ambiguities are removed, in contrast to earlier work.
\end{abstract}
\section{Introduction}\label{unu} 
In sharp contrast with the zero magnetic field case, the analysis of
properties of electrons in
periodic or random potentials 
subjected to external magnetic fields is a
very challenging problem. The difficulty is rooted in the singular
nature of the magnetic interaction: due to a linear increase of the
magnetic vector potential, the naive perturbation theory breaks down
even at arbitrarily small fields. 

To our best knowledge, only the periodic case has been
considered in connection with the Faraday effect for bulk systems. The first  full 
scale quantum computation was done by Laura 
M. Roth \cite{R2} (for a review of earlier attempts we direct the
reader to this
paper). The physical experiment starts by sending a monochromatic light wave,
parallel to the $0z$ direction and 
linearly polarized in the plane $x0z$. When the light enters the
material, the polarization plane can change; in fact, there exists a 
linear relation between the angle
$\theta$ of rotation of the plane of polarization per unit length and
the transverse component of the conductivity tensor $\sigma_{xy}$ 
(see formula (1)
in \cite{R2}). The material is chosen in such a way that when the
magnetic field is zero, this transverse component vanishes. When the
magnetic field $B$ is turned on, 
the transverse component is
no longer zero. For weak fields one expands the conductivity tensor to 
first order in $B$ and obtains a formula for the Verdet
constant. 

Therefore the central object is $\sigma_{xy}(B)$, which depends among
other things on temperature, density of the material, and 
frequency of light. 
Using a modified Bloch representation, Roth was able to obtain a
formula for $\frac{d\sigma_{xy}}{dB}(0)$, and studied how this first order term
behaves as a function of frequency, both for metals and
semiconductors. 

However, the theory in \cite{R2} is not free of
difficulties. First, it seems almost hopeless to estimate errors or to push
the computation to higher orders in $B$. Second, even the first order
formula contains terms which are singular at the crossings of the
Bloch bands. Accordingly, at the practical level this theory only met a
moderate success and alternative formalisms have been used, as for
example the
celebrated Kohn-Luttinger effective many band Hamiltonian (see
\cite{Jim, Ro, Pid} and references therein), or tight-binding models
 \cite{Pe}. Since all these methods have limited applicability, 
a more flexible approach was still needed. 

In the zero magnetic field case, a very successful formalism (see
e.g. \cite{Per, Huh, Ban, But} and references
therein) is to use the Green function method. This is based on the
fact that the traces involved in computing various physical quantities
can be written as integrals involving Green functions. 
The main aim of our paper is to develop a Green function approach to 
the Faraday effect, i.e. for the conductivity tensor when a magnetic
field is present. Let us point out that the use of Green functions 
(albeit  different from the ones used below) goes back at least to
Sondheimer and Wilson \cite{SW} in their theory of diamagnetism of
Bloch electrons. Aside from the fact that the Green function (i.e. the
integral kernel of the resolvent or the semi-group) is easier to
compute and control, the main point is that by factorizing out the so called
"non integrable phase factor" (or "magnetic holonomy") from the Green
function, one can cope with the singularities introduced by the increase at
infinity of the magnetic vector potentials. 
In addition, (as it has already been observed by Schwinger \cite{Sch} in
a QED context) after factorizing out the magnetic holonomy one remains
with a gauge invariant quantity which makes the problem of gauge
fixing irrelevant. The observation (going back at least to Peierls
\cite{Pei}) that one can use these magnetic phases in order to control
the singularity of the magnetic perturbation has been used many times
in various contexts (see e.g. \cite{SW, Lu}). We highlight here
the results of Nedoluha \cite{Ne} where a Green function approach for the
magneto-optical phenomena at zero temperature and with the Fermi level
in a gap has been investigated. 

But the power of this method has only
recently been fully exploited in \cite{CN, Cor}, and developed as a
general gauge invariant magnetic perturbation theory in \cite{N2}. 
Applied to the case at hand, this theory gives an expansion of the conductivity tensor in terms of the
zero magnetic field Green functions. Moreover, it is free of any
divergences. A key ingredient in controlling divergences is the
exponential decrease of the Green functions with the distance between
the arguments, for energies outside the spectrum \cite{Com, N1}. 
We stress the fact that since no basis is involved, periodicity is not
needed and the theory can also be applied to random systems. Finite
systems and/or special geometries (layers) are also allowed. The content of the paper is as follows: 

In Section \ref{sectiuneadoi} we give a
derivation of the conductivity tensor from first principles in the
linear response theory. We
include it to point out that it coincides with various formulas used
before. Although in physics establishing the Kubo formula is considered somehow a
triviality, from a mathematical point of view it remains a serious chalenge (see
\cite{BGKS}).  

Section \ref{sectiuneatrei} contains the precise formulation
of the thermodynamic limit, stated in Theorem \ref{th_limterm}. We do
not give its proof here, but we try to explain why it is true. 

Section \ref{sectiuneapatru} shows that at the
thermodynamic limit, at zero temperature, zero frequency, and for
the Fermi energy in a spectral gap, we re-obtain 
a formula of Streda \cite{Str} for the transverse component of the
conductivity tensor, known from the Integral
Quantum Hall Effect (IQHE). A precise statement and its proof are contained in
Theorem \ref{stredawidom}. 
Moreover, under the proviso
that exponentially localized Wannier function 
exist (see Theorem \ref{wanierx}), 
this transverse component vanishes (see also \cite{Th}, \cite{BN}
for related results). We stress that this result holds for the {\it
  whole} $\sigma_{xy}(B)$ as long as the magnetic field is not too
large, not just for $\frac{d\sigma_{xy}}{dB}(0)$. 
The vanishing of its first order
correction was in fact claimed in formula (50) in \cite{R2}.  

Section \ref{sectiuneacinci} contains the exact quantum computation 
of $\sigma_{xy}(B)$ for free electrons; 
in spite of the fact that such
a result might be known (and it is known at zero frequency), 
we were not able to find it in the literature. Interestingly enough, the
quantum computation gives the same result
as the well known classical computation (when the relaxation time 
is infinite).  

Section \ref{sectiuneasapte} contains the core of the paper, 
which includes the derivation of $\frac{d\sigma_{xy}}{dB}(0)$ for
general Bloch electrons. As in the zero magnetic field case, its formula
only contains zero magnetic field Green functions and current
operators. 

Section \ref{sectiuneaopt} deals with periodic systems, and the result of
the previous section is written down in terms of zero magnetic field
Bloch functions and bands.

At the end we have some conclusions.

The main goal of this
paper is to present the strategy, state the results concerning the
Verdet constant, and to outline future theoretical and
practical problems. Detailed proofs of the thermodynamic limit and of 
other technical estimates will be given elsewhere.

\section{Preliminaries. The conductivity tensor in the linear response
  regime} \label{sectiuneadoi}

We begin by fixing the notation used in the description of
independent electrons subjected to a constant magnetic field.
The units are chosen so that $\hbar
=1$. Since we consider spin $1/2$ particles, the one particle Hilbert
space for a non-confined particle is
$$
\mathcal{H}_{\infty}=L^2(\R^3)\oplus L^2(\R^3)
$$
with the standard scalar product. Accordingly, all operators below and
their integral kernels are $2\times 2$ matrices in the spin variable.
We choose the constant magnetic field of strength $B$ to be oriented
along the $z$-axis. Then the one particle Hamiltonian with the
spin-orbit coupling included is (see e.g. \cite{R2})
\begin{equation}\label{hamimare}
H_\infty(B)=\frac{1}{2m}{\bf P}(B)^2 +V+ g\mu_b B\sigma_{3},
\end{equation}
 with
\begin{equation}\label{impuls}
{\bf P}(B)=-i\nabla -b{\bf a}+
\frac{1}{2mc^2}{\bf s}\wedge (\nabla V)={\bf P}(0)-b{\bf a}
\end{equation}
where 
$$
b=-\frac{e}{c}B
$$
and ${\bf a}(\x)$ is an
arbitrary smooth magnetic vector potential which generates a magnetic field
of intensity $B=1$ i.e. $\nabla \wedge {\bf a}(\x)=(0,0,1)$. The most
frequently used magnetic vector potential is the symmetric gauge:
\begin{equation}\label{simg}
{\bf a}_{0}(\x)=\frac{1}{2}\n_{3}\wedge \x.
\end{equation}
where $\n_{3}$ is the unit vector along $z$ axis.

 In the periodic case we denote by
$\mathcal{L}$ the underlying Bravais lattice, by $\Omega$ its
 elementary cell and by $\Omega^*$ the corresponding Brillouin zone.  
$|\Omega|$ and  $|\Omega^*|$ stand for the volumes of the elementary cell and
Brillouin zone respectively. 
In the absence of the magnetic field one has the well known Bloch
representation in terms of Bloch functions:
\begin{equation}\label{blochf}
\Psi_j(\x,\bk)=\frac{1}{\sqrt{|\Omega^*|}}e^{i \bk\cdot\x}
u_j(\x,\bk),\quad \x\in\R^3
\end{equation}
where $u_j(\x,\bk)$ are the normalized to one eigenfunctions of the operator
\begin{equation}
h(\bk)u_j(\x,\bk)=\lambda_{j}(\bk)u_j(\x,\bk)
\end{equation}
\begin{align} \label{fibHa}
h(\bk) &= \frac{1}{2m}\left (-i\nabla_p +\frac{1}{2mc^2}{\bf s}\wedge (\nabla
V)+\bk\right )^2
+V,\nonumber \\
&= \frac{1}{2m}(\p+\bk)^2+V,  \quad \bk \in \Omega^* ,\\
\p &= -i\nabla_p +\frac{1}{2mc^2}{\bf s}\wedge (\nabla
V),\nonumber 
\end{align}
defined in  $L^2(\Omega)\oplus L^2(\Omega)$ with periodic boundary
conditions. We label $\lambda_{j}(\bk)$ in increasing order. We
have to remember that, as functions of $\bk$, $\lambda_{j}(\bk)$ and
 $u_j(\x,\bk)$ are not differentiable at the crossing points. Since
the $\Psi_j(\x,\bk)$'s form a basis of generalized eigenfunctions, the Green 
function (i.e. the integral kernel of the resolvent) writes as:
 \begin{equation}\label{intker}
G_{\infty}^{(0)}(\x,\y; z)=\int_{\Omega^*}\sum_{j\geq 1}
\frac{|\Psi_j(\x,\bk)  \rangle \langle \Psi_{j}(\y,\bk)| }
{\lambda_j(\bk)-z}d\bk, 
\end{equation}
and it is seen as a matrix in the spin variables. 
The above formula has to be understood in the formal sense since the series in
the right hand side is typically not absolutely convergent, and care
is to be taken
when interchanging  the sum with the integral. Notice however that $G_\infty^{(0)}(\x,\x'; z)$ is a well behaved
matrix valued function.

We consider a system of noninteracting electrons in the grand-canonical
ensemble. More precisely, we consider a box $\Lambda_1\subset \R^3$,
which contains the origin, and a family of scaled boxes
\begin{equation}\label{cutiacu gaz}
\Lambda_L=\{\x\in\R^3:\; \x/L\in \Lambda_1\}.
\end{equation}
The thermodynamic limit will mean $L\rightarrow \infty$, that is
when $\Lambda_L$ tends to fill out the whole space. The one particle Hilbert space is 
$\mathcal{H}_L:=L^2(\Lambda_L)\oplus L^2(\Lambda_L)$. The one
particle Hamiltonian is denoted by $H_L(B)$ and is given by 
(\ref{hamimare}) with 
Dirichlet boundary conditions (i.e. the wave-functions in the domain of
$H_L(B)$ vanish at the surface $\partial\Lambda_L$). More precisely,
we first define it on 
$C_0^\infty(\Lambda_L)\oplus C_0^\infty(\Lambda_L)$, and then 
$H_L(B)$ will be the Friedrichs extension of this minimal operator. This is indeed
possible, because our operator can be written as (up to some
irrelevant constants) 
$ -\Delta_D I_2+W,$ where $\Delta_D$ is the Dirichlet Laplacian and $W$
is a first order differential operator, relatively bounded
to $-\Delta_D I_2$ (remember that $L<\infty$) with relative bound zero.  The form domain of
$H_L(B)$ is the Sobolev space $H_0^1(\Lambda_L)\oplus H_0^1(\Lambda_L)$, while the operator
domain is 

\begin{equation}\label{domeniuop}
{\rm Dom}(H_L(B))=D_L\oplus D_L,\quad D_L:= H^2(\Lambda_L)\cap H_0^1(\Lambda_L).
\end{equation}
Moreover,
$H_L(B)$ is essentially self-adjoint on 
$C_{(0)}^\infty (\overline{\Lambda_L})$, i.e. functions with support in
$\overline{\Lambda_L}$ and indefinitely differentiable in $\Lambda_L$
up to the boundary.

We assume that the temperature $T=1/(k \beta)$ and the chemical potential $\mu$
are fixed by a reservoir of energy and particles. We work in a
second quantized setting with an antisymmetric Fock
space denoted by $\mathcal{F}_L$. Denote the operators in the Fock
space with a hat and borrow some notation from
the book of Bratelli and Robinson \cite{BR}: if $A$ 
is an operator defined in
$\mathcal{H}_L$, we denote by $ \hat{A}=d\Gamma(A)$ its second 
quantization in the Fock space. At $t=-\infty$ the system is
supposed to be in the grand-canonical equilibrium state 
of  temperature $T$ and 
chemical potential $\mu$, 
i.e. the density matrix is
\begin{equation}\label{matridensi}
\hat{\rho}_e= \frac{1}{{\rm Tr}(e^{-\beta \hat{K}_{\mu}})}{e^{-\beta  \hat{K}_{\mu}}},
\end{equation}
where
\begin{equation}\label{camiu}
\hat{K}_{\mu}= d\Gamma(H_L(B)-\mu\cdot {\rm Id})
\end{equation}
is the ``grand-canonical Hamiltonian''.

The interaction with a classical electromagnetic field is described by
a time dependent electric potential
\begin{equation}\label{eleccimp}
V(\x,t):=(e^{i\omega t}+e^{-i\overline{\omega} t})e{\bf E}\cdot\x,\;
t\leq 0,\; \x\in \Lambda_L.
\end{equation} 
so the total time dependent one-particle Hamiltonian is 
\begin{equation}
H(t)= H_L(B)+V(t).
\end{equation}
 Notice that $e$ near ${\bf E}$ is the
positive elementary charge. Here we take ${\rm Im}\;\omega <0$ which 
plays the role of an adiabatic parameter, and insures that there is
no interaction in the remote past.  
Finally, the one-particle current operator is as usual
\begin{equation}\label{curent1}
{\bf J}=-e i [H_L(B), {\bf X}]=-\frac{e}{m}{\bf P}(B),
\end{equation}
where ${\bf X}$ is the multiplication by $\x$. Note that {\bf J} is a
well defined operator on the domain of $H_L(B)$, because
multiplication by any component of ${\bf X}$ leaves this domain
invariant (see \eqref{domeniuop}). Moreover, since $L<\infty$,  ${\bf
  X}$ is a bounded operator. In fact, ${\bf X}$ is the true physical
self-adjoint observable, while ${\bf P}(B)$ (or ${\bf J}$) appear when one
differentiates the map $t\mapsto e^{itH_L(B)} ${\bf X}$e^{-itH_L(B)}$
in the strong sense on the domain of $H_L(B)$.

We assume that the state of our system is now described
by a time-dependent density matrix, $\hat{\rho}(t)$, obtained by
evolving $\hat{\rho}_e$ from $-\infty$ up to the given time, i.e.
\begin{equation}\label{liouville}
i\partial_t \hat{\rho}(t)=[\hat{H}(t),\hat{\rho}(t)],\quad \hat{\rho}(-\infty)=\hat{\rho}_e.
\end{equation}
Going to the interaction picture and using the Dyson expansion up to
the first order, one gets
\begin{equation}\label{linresp}
\hat{\rho}(t=0)=\hat{\rho}_e -i
\int_{-\infty}^0[d\Gamma(\tilde{V}(s),\hat{\rho}_e]ds+{\cal O}({\bf E}^2),
\end{equation}
where
\begin{equation}\label{interdet}
{\tilde V}(s):=e^{isH_L(B)} V(s)e^{-isH_L(B)}. 
 \end{equation}
The current density flowing through 
our system at $t=0$ is given by (see (\ref{linresp})):
\begin{align}\label{curent2}
{\bf j} &= \frac{1}{|\Lambda_L|}{\rm Tr}_{\mathcal{F}_L}\left (\hat{\rho}(0) \hat{{\bf
    J}}\right )=\frac{1}{|\Lambda_L|}{\rm Tr}_{\mathcal{F}_L}\left (\hat{\rho}_e  \hat{{\bf
    J}}\right )\nonumber \\
&- \frac{i}{|\Lambda_L|}{\rm Tr}_{\mathcal{F}_L}\left (\int_{-\infty}^0
    [d\Gamma(\tilde{V}(s)),\hat{\rho}_e]  \hat{{\bf
    J}}ds \right
    )+{\cal O}({\bf E}^2).
\end{align}
In evaluating the r.h.s. of (\ref{curent2}) we use the well known
fact that traces over the Fock space can be
computed in the one-particle space (see Proposition 5.2.23 in \cite{BR}):
\begin{equation}\label{com3}
{\rm Tr}_{{\cal F}_L}\left \{\hat{\rho}_e d\Gamma(A)\right \}=
{\rm Tr}_{{\cal H}_L}\left \{f_{FD}(H_L(B)) { A}\right \},
\end{equation}
where $f_{FD}$ is the Fermi-Dirac one-particle distribution function:
\begin{equation}\label{fermidira}
f_{FD}(x):=\frac{1}{e^{\beta(x-\mu)}+1},\quad x\in \R, \beta >0,\mu\in
\R. 
\end{equation}
Plugging (\ref{com3}) into (\ref{curent2}), the identity
$[d\Gamma(A),d\Gamma(B)]=d\Gamma([A,B])$, the invariance of trace
under cyclic permutations and ignoring the  quadratic correction in
${\bf E}$ one arrives at
\begin{align}\label{curent3}
{\bf j} &=\frac{1}{|\Lambda_L|} {\rm Tr}_{\mathcal{H}_L}\left \{f_{FD}(H_L(B)) {\bf
    J}\right \}\nonumber \\
& -\frac{i}{|\Lambda_L|}\frac{e}{m}{\rm Tr}_{\mathcal{H}_L}\left (\int_{-\infty}^0
    [\tilde{V}(s),{\bf P}(B)] f_{FD}(H_L(B))ds\right
    ).
\end{align} 
The first term in (\ref{curent3}) is always zero because of the
identity (trace cyclicity again)
\begin{equation}\label{cyklyc}
{\rm Tr}_{\mathcal{H}_L}\left \{
[H_L(B), {\bf X}]f_{FD}(H_L(B)) \right \}={\rm Tr}_{\mathcal{H}_L}\left
\{[f_{FD}(H_L(B)),H_L(B)] {\bf X}\right \}=0.
\end{equation}
which is nothing but the fact that the current vanishes on an 
equilibrium state. Note that these operations under the trace sign are
quite delicate, since unbounded operators are involved. Let us for
once give a complete proof to \eqref{cyklyc}. We have the identity
between bounded operators
(consider the first component $X_1$):
\begin{equation}\label{cyklycmat}
[H_L(B), X_1]f_{FD}(H_L(B))=H_L(B) X_1 f_{FD}(H_L(B))-X_1H_L(B)f_{FD}(H_L(B)).
\end{equation}
Remember that $X_1$ is a bounded operator in the box, and preserves
the domain of $H_L(B)$. This 
means that
the operator $O_L=(H_L(B)+i)X_1 (H_L(B)+i)^{-1}$ is bounded. Hence we
can write 
$$H_L(B) X_1 f_{FD}(H_L(B))=[1-i(H_L(B)+i)^{-1}]\;O_L \; [H_L(B)+i]f_{FD}(H_L(B)).$$ 
Now the operator $[H_L(B)+i]f_{FD}(H_L(B))$ still is trace class due to the
exponential decay of $f_{FD}$, while $[1-i(H_L(B)+i)^{-1}]$ and $O_L$
are bounded. Thus  $H_L(B) X_1 f_{FD}(H_L(B))$ is trace class and we
can compute its trace using the complete eigenbasis of $H_L(B)$, which
gives the same result as for the other operator $X_1 H_L(B)
f_{FD}(H_L(B))$. Thus \eqref{cyklyc} is proved.

Using (\ref{eleccimp}) and (\ref{interdet}) one can write
\begin{eqnarray}\label{curent5}
j_\alpha = \sum_{\beta =1}^3 
\{\sigma_{\alpha\beta}(\omega)+\sigma_{\alpha\beta}(-\overline{\omega})\}
E_\beta, \quad \alpha\in \{1,2,3\},\; \Im(\omega)<0, 
\end{eqnarray}
where the conductivity tensor is given by 
\begin{align}\label{curent6}
& \sigma_{\alpha\beta}(B, \omega)=\\
&-\frac{i}{|\Lambda_L|}\frac{e^2}{m}{\rm Tr}_{\mathcal{H}_L}\int_{-\infty}^0
    [e^{is H_L(B)}x_\beta e^{-is H_L(B)}, P_\alpha(B)] 
f_{FD}(H_L(B))e^{is\omega}ds.\nonumber 
\end{align}
Performing an integration by parts, using the formulas $i[H_L(B),x_\beta]=P_\beta (B)
/m$ and $i[P_\alpha (B),x_\beta]=\delta_{\alpha\beta}$ one
arrives at \begin{eqnarray}\label{curent8}
&{}&\sigma_{\alpha\beta}(B,\omega)=
\frac{1}{|\Lambda_L|}\frac{ e^2}{i m\omega }\{ \delta_{\alpha\beta}{\rm
  Tr}(f_{FD}(H_L(B))) \\
&+& \frac{i }{m }{\rm Tr}\int_{-\infty}^0
 e^{is(\omega + H_L(B))}P_\beta(B) e^{-is H_L(B)}
[P_\alpha(B),f_{FD}(H_L(B))] ds \},\nonumber 
\end{eqnarray}
and this coincides (at least at the formal level)  with  formula (5)
in \cite{R2}. Notice that from now on, we write just Tr when we perform the trace,
since we only work in the one-particle space.

Since we are interested in the Faraday effect, and we assume that the
magnetic field ${\bf B}$ is parallel with the $z$ axis, we will only consider
the transverse conductivity $\sigma_{12}(B,\omega)$. Hence the
first term vanishes. 
We now perform the integral over $s$ with the help of Stone's formula
followed by a deformation of the contour (paying attention not to hit
the singularities of $f_{FD}(z)$ or to make the integral over $s$ divergent
\begin{equation}\label{stone1}
f_{FD}(H_L(B))
e^{is (H_L(B)+ \eta)}=\frac{i}{2\pi}\int_{\Gamma_\omega}
f_{FD}(z)e^{is (z+\eta)}(H_L(B)-z)^{-1}dz.
\end{equation}
where $\eta$ is either $0$ or $\omega$, the contour is
counter-clockwise oriented and given by
\begin{equation}\label{contur1}
\Gamma_\omega=\left \{x\pm id:\;a\leq x
  <\infty
\right \}\bigcup 
\left \{a+i y:\:-d\leq y \leq
d\right \} 
\end{equation}
with 
\begin{equation}\label{bo}
d=\min\left\{ \frac{\pi}{2\beta},\frac{|{\rm Im}\;\omega|}{2}\right\},
\end{equation}
and $a+1$ lies below the spectrum of $H_L(B)$. As a final result one gets
\begin{align}\label{stone5}
&  \sigma_{12}(B,\omega)= -\frac{e^{2}}{2\pi m^{2}\omega |\Lambda_L|} \\
&\cdot {\rm Tr}
\int_{\Gamma_\omega}
{f}_{FD}(z) 
\left \{ P_1(B)(H_L(B)-z)^{-1}P_2(B) 
(H_L(B)-z-\omega)^{-1}\right .\nonumber \\
&+ \left . z \rightarrow z-\omega \right \}dz=:
\frac{e^{2}}{ m^{2}\omega}a_L(B,\omega)
\nonumber 
\end{align}
where ``$z \rightarrow z-\omega $'' means a similar term where 
we exchange $z$ with $z-\omega$. Now one
 can see that by inserting the eigenbasis of $H_L(B)$ one
 obtains the well known formula derived from semi-classical radiation
 theory (see e.g. formula (4) in \cite{R2}).

\section{Gauge invariance and existence of the thermodynamic limit}\label{sectiuneatrei}

Up to now the system was confined in a box $\Lambda_{L}$. As is well
known (see e.g. \cite{R2}) a direct evaluation of (\ref{stone5}) (or
previous formulas equivalent to it including formula (4) in Roth's
paper) is out of reach: the eigenvalues and eigenstates of $H(B)$ are
rather complicated (even in the thermodynamic limit
$\Lambda_{L}\rightarrow \R^{3}$) and at the same time the Bloch
representation is plagued by singular matrix elements of the magnetic
vector potential. Roth used a modified magnetic Bloch representation in
\cite{R1} and derived a formula for the linear term in $B$ of
(\ref{stone5}) 
in terms of the zero magnetic field Bloch representation. Still, her
procedure is not free of difficulties since it involves
$\nabla_{\bk}u_{j}(\x,\bk)$ which might not exist at crossing points. In
addition, it seems almost hopeless to control the errors or to push
computations to the second order in $B$ which would describe the
Cotton-Mouton effect for example.

In what follows, we shall outline another route of evaluating
(\ref{stone5}) which is mathematically correct, systematic, and 
completely free of the above
difficulties. There are two basic ideas involved. The
first one (going back at least to Sondheimer and Wilson \cite{SW} in their
theory of diamagnetism) consists in writing the trace in 
(\ref{stone5}) as integrals over  $\Lambda_{L}$ of corresponding
integral kernels. This is nothing but the well known Green function
approach (see e.g. \cite{Gon}) which has been 
very successful in computing optical and magneto-optical properties of
solids (see e.g. \cite{Per}, \cite{Huh}, \cite{Ban}) 
in the absence of an external magnetic field.
The point  is that the integral kernels are on one hand
easier to control and compute,  and on the other hand they do not
require periodicity. Moreover, this approach proved to be 
essential in deriving rigorous results concerning the diamagnetism of free electrons
\cite{Cor, ABN} and actually we expect the methods of
the present paper to simplify the theory of diamagnetism of Bloch electrons
as well.

However, when applying Green function approach in the presence of an external magnetic field one hits again the
divergences caused by the linear increase of the magnetic vector
potential: naively, at the first sight $a_L(B,\omega)$ is not bounded
in the thermodynamic limit $L \rightarrow \infty$  but instead grows like the second power of
$L$.
It was already observed in \cite{ABN} that these divergent terms
vanish identically due to some identities coming from gauge
invariance. 

This is indeed the case and the main point of this paper is to show,
following the developments in  \cite{CN}, \cite{Cor}, \cite{N2}, that factorizing
the so called `` non-integrable phase factor'' from the Green 
function (the integral kernel of $(H_L(B)-\zeta)^{-1}$) allows, at the same
time, to eliminate the divergences coming from the increase of the
magnetic vector potential and to obtain a controlled expansion in
powers of $B$. In addition, this leads to expressions of $a_L(B,\omega)$
which are manifestly gauge invariant.

For an arbitrary pair of points $\x$, $\y\in \Lambda_{L}$  consider the
``magnetic phase'' associated with the magnetic vector potential
$\ba(\bf u)$ defined as the path integral 
on the line linking $\y$ and $\x$:
\begin{equation}\label{intkerb2}
\phi_{\ba}(\x,\y)=\int_{\y}^{\x}\ba({\bf{u}})\cdot d{\bf u}. 
\end{equation}
The magnetic phase satisfies the following crucial identity: for every
fixed $\bc$
\begin{align}\label{identit1}
e^{-ib \phi_{\ba}(\x ,\bc)}\mathbf{P}(B)e^{i b\phi_{\ba}
(\x,\bc 
)}=\mathbf{P}(0)-b {\bf A}(\x-\bc). 
\end{align}
where $
{\bf A}(\x)=\frac{1}{2}
{\bf n}_3\wedge
\x $, 
i.e. irrespective of the choice of $\ba(\x)$, 
\begin{equation}\label{ident1}
{\bf A}(\x-\bc)=\frac{1}{2}
{\bf n}_3\wedge
(\x-\bc)  
\end{equation}
is the symmetric (transverse, Poincar\'e) gauge with respect to $\bc$.

Write now the Green function (as a $2\times 2$ matrix in the spin space)
\begin{equation}\label{green}
G_{L}(\x,\y;\zeta)=(H_{L}-\zeta)^{-1}(\x,\y)
\end{equation}
in the factorized form
\begin{equation}\label{gki}
G_{L}(\x,\y;\zeta)=e^{i b\phi_{\ba}
(\x,\y 
)}K_{L}(\x,\y;\zeta).
\end{equation}
It is easy to check that while
$G_{L}(\x,\y;\zeta)$ is gauge dependent,
$K_{L}(\x,\y;\zeta)$ is gauge independent i.e. the whole
gauge dependence of $G_{L}(\x,\y;\zeta)$ is contained in the
phase factor $e^{i b\phi_{\ba}
(\x,\y 
)}$. Plugging the factorization (\ref{gki}) into the integrand of the
r.h.s. of (\ref{stone5}), using (\ref{identit1}) and (\ref{ident1}), one
obtains that its integral kernel writes as
\begin{align}\label{calA}
& \mathcal{A}_{s,s'}^L(\x,\x')=
e^{i b\phi_{\ba}
(\x,\x' 
)} \\
& \cdot \int_{\Gamma_\omega}dz{f}_{FD}(z) 
\sum_{\sigma=1}^2\int_{\Lambda_{L}}d\y e^{ib\Phi(\x,\y,\x')}
\{[(P_{1,\x}(0)-b
  A_1(\x-\y))K_{L}(\x,\y;z)]_{s,\sigma}\nonumber \\
&\cdot [(P_{2,\y}(0)-bA_2(\x-\y))K_{L}(\y,\x';z+\omega)]_{\sigma, s'}+z\rightarrow
z-\omega\},\nonumber 
\end{align}
where $$
\Phi(\x,\y,\x')=\phi_{\ba}(\x ,\y)+\phi_{\ba}(\y ,\x')+\phi_{\ba}(\x' ,\x)
$$
 is the flux of the magnetic field $(0,0,1)$
through the triangle $\Delta(\x,\y,\x')$. Now the fact that there are no
long range divergences in the formula for
$\mathcal{A}_{s,s'}(\x,\x')$ follows from the exponential decay of
Green functions \cite{Com} (see also \cite{N1}): for $\zeta$ outside the
spectrum of $H$ there exists $m(\zeta)>0$ such that as
$|\x-\y|\rightarrow \infty$
\begin{equation}\label{expocade}
|K_{L}(\x,\y;\zeta)|=|G_{L}(\x,\y;\zeta)|\sim
e^{-m(\zeta)|\x-\y|}.
\end{equation}
It can be proved (the technical details which are far from being
simple will be given elsewhere) that $\mathcal{A}_{s,s'}^L(\x,\x')$ is
jointly continuous and moreover outside a thin region near the surface
of  $\Lambda_{L}$ one can replace it by the integral kernel 
 $\mathcal{A}^{\infty}_{s,s'}(\x,\x')$ of the corresponding operator
 on the whole $\R^{3}$. Accordingly, up to surface corrections:
\begin{equation}\label{ik2}
a_L(B,\omega )\approx -\frac{1}{2\pi|\Lambda_{L}|}\sum_{s=1}^2\int_{\Lambda_{L}}
\mathcal{A}^{\infty}_{s,s}(\x,\x)d\x.
\end{equation}
Notice that due to the fact  that
$\Phi(\x,\y,\x)=\phi_{\ba}(\x ,\x)=0$ the phase factors appearing in (\ref{calA}) reduce to unity in (\ref{ik2}).

In the periodic case, from the fact that in the symmetric gauge the
Hamiltonian $H_{\infty }(B)$ commutes with the magnetic translations
(actually one can define magnetic translations for 
an arbitrary gauge, just first make the gauge transformation relating $\ba (\x )$ to ${\bf A} (\x )$)
generated by
$\mathcal{L}$, it follows that for
$\ggamma \in \mathcal{L}$ we have:
$$
K_{\infty}(\x+\ggamma,\y+\ggamma;\zeta)=K_{\infty}(\x,\y;\zeta),
$$
which implies that
$$
\mathcal{A}^{\infty}_{s,s}(\x+\ggamma,\x+\ggamma)=\mathcal{A}^{\infty}_{s,s}(\x,\x)
$$
is periodic with respect to $\mathcal{L}$, 
hence up to surface corrections:
\begin{equation}\label{stone8}
a_L(B,\omega )\approx a(B,\omega )= -\frac{1}{2\pi |\Omega|} \sum_{s=1}^2
\int_\Omega \mathcal{A}^\infty_{s,s}(\x,\x)d\x.
\end{equation}
Therefore, the transverse conductivity writes as
\begin{equation}\label{condupevol}
\sigma_{12}(B,\omega )=\frac{e^{2}}{m^{2}\omega} a(B,\omega)
\end{equation}
with $a(B,\omega)$ given by the r.h.s. of (\ref{stone8}). 

A precise formulation of this result is contained in the following
theorem:

\begin{theorem}\label{th_limterm}
Assume for simplicity that $\Omega$ is the unit cube in $\R^3$. The above defined transverse component of the conductivity tensor admits the thermodynamic
limit; more precisely:

{\rm i.} The following operator defined by a $B(L^2\oplus L^2)$-norm convergent
Riemann integral,
\begin{align}\label{efel}
F_L&:=-\frac{1}{2\pi } \int_{\Gamma_\omega}{f}_{FD}(z) \{ P_1(B)(H_L(B)-z)^{-1}P_2(B) 
(H_L(B)-z-\omega)^{-1}\nonumber \\
&+ z \rightarrow z-\omega \}dz,
\end{align} 
is in fact trace-class, and
$\sigma_{12}^{(L)}(B,\omega)=\frac{e^{2}}{
  m^{2}\omega |\Lambda_L|}{\rm Tr}(F_L)$.  

{\rm ii.} Consider the operator $F_\infty$ defined by the same integral but
with $H_\infty(B)$ instead of $H_L(B)$, and defined on the whole
space. Then $F_\infty$ is an integral
operator, with a kernel $\mathcal{A}_{s,s'}^\infty(\x,\x')$ jointly
continuous on its spatial variables. Moreover, the function defined by 
$\R^3\ni \x\to s_{B}(\x):=\sum_{s=1}^2\mathcal{A}_{s,s}^\infty(\x,\x)\in \mathbb{C}$
is continuous and periodic with respect to $\mathbb{Z}^3$. 

{\rm iii.}  The thermodynamic limit exists: 
\begin{align}\label{efel22}
\sigma_{12}^{(\infty)}(B,\omega):=\lim_{L\to
  \infty}\sigma_{12}^{(L)}(B,\omega)=\frac{e^{2}}{m^{2}\omega |\Omega|}\int_\Omega s_{B}(\x) d\x.
\end{align}
\end{theorem}

\vspace{0.5cm}

The proof of this theorem will be given elsewhere \cite{CN2}.

\section{The zero frequency limit at $T=0$: a rigorous proof  
of  the Widom-Streda formula for semiconductors}\label{sectiuneapatru}

Doing some very formal computations one can show that at $T=0$ and
$\omega=0$,  $\sigma_{12}(B,\omega)$ as given by (\ref{stone8}) and
(\ref{condupevol}) coincide with the formula for the quantized Hall
conductivity (see e.g. formulas (5),(6) in \cite{Str}) which in turn
gives (again at the heuristic level) the well known Widom-Streda
formula. The original derivation has little mathematical rigor, in
particular because it assumes some very strong assumptions on the
existence and regularity of $(H_\infty(B)-\lambda +i 0)^{-1}$ as a
function of $\lambda\in\mathbb{R}$. These assumptions are clearly not
true in many situations. 

Here we will show how the Widom-Streda formula can be rigorously obtained when
the Fermi energy lies in a spectral gap. The problem in which the Fermi
energy is in the spectrum, remains open. 
Now assume that for some $B$,  the chemical potential $\mu$ lies in a spectral
gap of $H_\infty(B)$. More precisely, throughout this section 
we suppose that $(d_1,d_2)\subset \rho (H_{\infty}(B))$ with $d_1<d_2$, 
and take $\mu\in (d_1,d_2)$. For simplicity, assume that $\mu
=\frac{d_1+d_2}{2}$. This is the typical situation for semiconductors
and/or isolators. In the absence of spin, 
the Widom-Streda formula roughly states that:

\begin{equation}\label{widomstreda}
\sigma_{12}(B,T=0,\omega=0)=ec\left. \frac{\partial N(B,E)}{\partial B}\right 
|_{E=\mu},
\end{equation}
where $N(B,E)$ is the integrated density of states up to the energy
$E$. When the spin is present (this was not considered by Streda),
this formula is slightly changed. If we denote by $B_1$ the $B$ 
multiplying the spin matrix
$\sigma_3$ in our Hamiltonian (\ref{hamimare}), and with $B_2$ the $B$ near
$\mathbf{A}$,  then in fact we have
\begin{equation}\label{widomstreda20}
\sigma_{12}(B,T=0,\omega=0)=ec\left. \frac{\partial N(B_1,B_2,E)}{\partial B_2}\right 
|_{E=\mu,\; B_1=B_2=B}.
\end{equation}

In the rest of this section we give a rigorous (but still not fully
technical) proof of (\ref{widomstreda20}). 

\begin{theorem}\label{stredawidom} Consider the conductivity at the thermodynamic limit 
given in \eqref{efel22}, and drop the superscript $\infty$. Then if we
first take the limit $T\searrow 0$, and after that $\omega\to 0$, we get: 
\begin{equation}\label{widomstredar}
\lim_{\omega \rightarrow 0}\lim_{T \rightarrow 0}
\sigma_{12}(B,T,\omega)=e c \frac{\partial}{\partial B_{2}}
\left . \frac{1}{ |\Omega|} \sum_{s=1}^2
\int_\Omega \Pi^{B}_{s,s}(\x,\x)d\x \right 
|_{B_1=B_2=B}, 
\end{equation}
where
\begin{equation}\label{proj}
\Pi^{B}=
\frac{i}{2\pi}\int_\Gamma\frac{1}{H_{\infty}(B)-z}dz
\end{equation}
 with a positively oriented contour $\Gamma$ 
enclosing the spectrum of $H_{\infty}(B)$
 below $\mu$, i.e. $\Pi^{B}$ is the Fermi projection onto the subspace
 of ``occupied'' states at $T=0$. 
\end{theorem}
\vspace{0.5cm} 
\noindent {\bf Remarks:}

\begin{enumerate}
\item  Streda did
not consider spin in his work \cite{Str} and in this case the derivative with respect to the magnetic field appears in the r.h.s. of (\ref{widomstredar}).

\item While it is not clear that  $\Pi^{B}(\x,\x)$ is well defined
($(H_{\infty}(B)-z)^{-1}(\x,\x)$ does not exist!) 
this can be seen by writing for some $a\in \rho(H_{\infty}(B))$:

\begin{align}\label{regularizarek}
\Pi^{B}&=
\frac{1}{2\pi}\int_\Gamma((H_{\infty}(B)-z)^{-1}-(H_{\infty}(B)-a)^{-1}dz\nonumber
\\
&=\frac{1}{2\pi}\int_\Gamma (z-a)(H_{\infty}(B)-z)^{-1}(H_{\infty}(B)-a)^{-1}dz.
\end{align}
Each resolvent has a polar integral kernel with a $1/|\x-\x'|$
singularity, and the product of two resolvents will have a continuous
kernel. In fact we can repeat this trick and obtain products of 
as many resolvents
as we want, thus further improving the regularity of the integral kernel.  
Technical details will be given elsewhere. 
Actually this kind of argument can 
be used to show that all operators defined by integrals over complex
contours have jointly continuous integral kernels.

\item Although  the order of limits in (\ref{widomstredar}) is
  important for the argument below, it might be possible 
(at least under additional conditions on the spectrum of 
$H_{\infty}(B)$) to interchange the order of limits. 
The important fact is that the thermodynamic limit has to be taken first: great care is to be taken when defining currents in the static limit for finite systems (for a discussion of this point in a related context see \cite{NSB}.

\item The result is valid for arbitrary magnetic field $B$ and
  establishes the connection between the Hall conductivity and the
  Faraday effect. However, the quantum Hall effect requires high 
magnetic fields while the Faraday effect is usually considered at 
low magnetic fields.
\end{enumerate}

\noindent {\bf Proof.} We start from the conductivity in the thermodynamic limit as given by Theorem \ref{th_limterm}:

\begin{equation}\label{stone6}
  \sigma_{12}(B,T,\omega)= -\frac{e^{2}}{2\pi m^{2}\omega |\Omega|}
\sum_{s=1}^{2} 
\int_{\Omega}\left [
\int_{\Gamma_{\beta,\omega}}
{f}_{FD}(z)\Sigma (z,\omega) 
dz \right ]_{s,s}(\x,\x) \; d\x,
\end{equation}
where
\begin{equation}\label{sigmadez}
\Sigma (z,\omega):=
 P_1(B)(H_\infty(B)-z)^{-1}P_2(B) 
(H_\infty(B)-z-\omega)^{-1} 
+  (z \rightarrow z-\omega).
\end{equation}

Since we made the assumption that  $(d_1,d_2)\subset\rho
 (H_{\infty}(B)$, then for $|\omega|<\frac{d_2-d_1}{4}$ the integral
 over $z$ on the contour $\Gamma_{\beta,\omega}$ 
can be replaced with the integral on the contour  
$\Gamma^{1}_{\beta,\omega} \cup  \Gamma^{2}_{\beta,\omega} $ where
 (see also \eqref{contur1} and \eqref{bo}):

\begin{align}
\Gamma^{1}_{\beta,\omega}&=\left \{x\pm id:\;a\leq x
  \leq d_1+\frac{d_2-d_1}{4}
\right \}\bigcup 
\left \{a+i y:\:-d\leq y \leq
d\right \}\nonumber \\ 
&\bigcup 
\left \{d_1+\frac{d_2-d_1}{4}+i y:\:-d\leq y \leq
d\right \} 
\end{align}
and 
\begin{align}
\Gamma^{2}_{\beta,\omega}&=\left \{x\pm id:\; x\geq d_2-\frac{d_2-d_1}{4}
\right \}\nonumber \\
&\bigcup 
\left \{d_2-\frac{d_2-d_1}{4}+i y:\:-d\leq y \leq
d \right \}. 
\end{align}
Accordingly, one can rewrite $\sigma_{12}(B,T,\omega)$ as
\begin{align}\label{stone7}
&  \sigma_{12}(B,T,\omega)= -\frac{e^{2}}{2\pi m^{2}\omega |\Omega| }
\sum_{s=1}^{2} 
\int_{\Omega}\\
&\left \{ 
\int_{\Gamma^{1}_{\beta,\omega}}\Sigma (z,\omega) 
dz\right .\nonumber \\
&\left . +\int_{\Gamma^{1}_{\beta,\omega}}(f_{FD}(z)-1)\Sigma (z,\omega)dz
+\int_{\Gamma^{2}_{\beta,\omega}}f_{FD}(z)\Sigma (z,\omega)dz
\right \}_{s,s}(\x,\x) \;d\x.
\nonumber 
\end{align}
Note that since the singularities of $f_{FD}(z)$ lie on $\frac{c+d}{2}+iy,\;y\in (-\infty, \infty)$, one can take $\Gamma^{j}_{\beta,\omega}$ independent of $\beta$ i.e. one can take $d=\frac{|\Im (\omega)|}{2}$ in (\ref{bo}).
At this point we take the limit $\beta \rightarrow \infty$. Since on 
 $\Gamma^{2}_{\omega}$ we have $|f_{FD}(z)| \leq 2 \exp[-\beta
 (x-\frac{d_1+d_2}{2})]$, 
and on  $\Gamma^{1}_{\omega}$ we have that  
 $|f_{FD}(z)-1| \leq 2 \exp[-\beta (\frac{d_1+d_2}{2}-x)]$,
the last two terms in (\ref{stone7}) vanish in the zero temperature
limit (full details about the control of various integral kernels will
be given elsewhere). Hence we get:
\begin{align}\label{stone77}
&  \sigma_{12}(B,T=0,\omega)= 
-\frac{e^{2}}{2\pi m^{2}\omega |\Omega| }
\sum_{s=1}^{2} 
\int_{\Omega} \left \{
\int_{\Gamma^{1}_{\omega}}\Sigma (z,\omega) 
dz 
\right \}_{s,s}(\x,\x) \;d\x.
\end{align}
An application of the Cauchy residue theorem shows that 
the two terms of  $\Sigma (z,\omega)$ (see \eqref{sigmadez}) will
combine in the above integral and give

\begin{align}\label{widomstredar1}
& \sigma_{12}(B,T=0,\omega)=
-\frac{e^{2}}{2\pi m^{2}\omega }
\frac{1}{ |\Omega|} \sum_{s=1}^2\\
&
\int_\Omega \left [\int_{\Gamma} 
P_1(B)(H_\infty(B)-z+\omega/2)^{-1}P_2(B) 
(H_\infty(B)-z-\omega/2)^{-1}\right ]_{s,s}(\x,\x)d\x.\nonumber
\end{align}
where $\Gamma$ is any finite contour such that $\Gamma  \subset
\rho(H_{\infty}(B)+\omega)$  for all $|\omega|<\frac{|d_2-d_1|}{4}$
and only enclosing the spectrum of $H_{\infty}(B)$ below
$\frac{d_1+d_2}{2}$. Now the integrand in (\ref{widomstredar1}) is
analytic in $\omega$ in a neighborhood of the origin. By expanding
the resolvents one obtains:

\begin{align}\label{widomstredar2}
&\sigma_{12}(B,T=0,\omega)=\\
&
-\frac{e^{2}}{2\pi m^{2}}
\frac{1}{ |\Omega|} \sum_{s=1}^2
\int_\Omega \left \{\int_{\Gamma} \frac{1}{\omega}P_1(B)(H_\infty(B)-z)^{-1}P_2(B) 
(H_\infty(B)-z)^{-1}\right .\nonumber \\ 
&+ \frac{1}{2}\left [ P_1(B)(H_\infty(B)-z)^{-1}P_2(B) 
(H_\infty(B)-z)^{-2}\right .  \nonumber\\
&\left . \left . -  P_1(B)(H_\infty(B)-z)^{-2}P_2(B) 
(H_\infty(B)-z)^{-1}\right ]\frac{}{}\right \}_{s,s}(\x,\x)d\x +\mathcal{O}(\omega ).\nonumber
\end{align} 
Apparently we have a first oder pole at $\omega=0$. But we now prove 
that the singular term in the r.h.s. of (\ref{widomstredar2}) is
identically zero. Namely (when no spin variables appear the integral 
kernels below have to be understood as matrices in the spin space):
\begin{equation}\label{nocontr}
\left \{\int_{\Gamma} P_1(B)(H_\infty(B)-z)^{-1}P_2(B) 
(H_\infty(B)-z)^{-1}dz\right \}(\x,\x)=0.
\end{equation}
Using the magnetic perturbation theory and the trick from
\eqref{regularizarek} one can prove that even though the integrand in
\eqref{nocontr} has a quite singular kernel, after integration with
respect to $z$ one gets a smooth kernel, exponentially localized near
the diagonal (details will be given elsewhere).   

Let us notice an operator equality which makes sense on compactly
supported functions: 
\begin{align}\label{nocontr100}
\frac{1}{m}(H_\infty(B)-z)^{-1}P_2(B) 
(H_\infty(B)-z)^{-1}=i [X_{2}, (H_\infty(B)-z)^{-1}],
\end{align}
because the resolvent $(H_\infty(B)-z)^{-1}$ sends compactly supported
functions into exponentially decaying functions (see
\eqref{expocade}), which are in the domain of $X_2$. In fact, the
operator on the right side has a nice integral kernel, given by 
\begin{align}\label{nocontr200}
\left \{i [X_{2}, (H_\infty(B)-z)^{-1}]\right\}(\x,\y)=i(x_2-y_2)G_\infty(\x,\y;z),
\end{align}
which is no longer singular at the diagonal and still exponentially
localized near the diagonal, thus defining a bounded operator on the
whole Hilbert space. After integration we get:
\begin{align}\label{nocontr1}
& \frac{i}{2\pi}\int_{\Gamma} P_1(B)(H_\infty(B)-z)^{-1}P_2(B) 
(H_\infty(B)-z)^{-1}dz=\\
&
\frac{i m}{2\pi}\int_{\Gamma} P_1(B)[(H_\infty(B)-z)^{-1},X_{2}]dz=m
 [P_1(B)\Pi ^{B},X_{2}],\nonumber
\end{align}
where we used that $P_1(B)$ and $X_2$ commute. Note that the magnetic
perturbation theory states that the integral kernel of $P_1(B)\Pi
^{B}$ is smooth and exponentially localized near the
diagonal. Therefore $[P_1(B)\Pi ^{B},X_{2}]$ will have the integral
kernel:
\begin{equation}\label{nocontr2}
\left\{ [P_1(B)\Pi ^{B},X_{2}]\right \}(\x,\y)=
(y_{2}-x_{2})\{P_1(B)\Pi ^{B}\}(\x,\y)
\end{equation}
which is identically zero at the diagonal and proves
(\ref{nocontr}). We can conclude at this point that:

\begin{align}\label{widomstredar20}
&\lim_{\omega\to 0}\sigma_{12}(B,T=0,\omega)=\\
&
-\frac{e^{2}}{4\pi m^{2}}
\frac{1}{ |\Omega|} \sum_{s=1}^2
\int_\Omega \left \{ \frac{}{}P_1(B)(H_\infty(B)-z)^{-1}P_2(B) 
(H_\infty(B)-z)^{-2}\right .  \nonumber\\
&\left . -  P_1(B)(H_\infty(B)-z)^{-2}P_2(B) 
(H_\infty(B)-z)^{-1}\frac{}{}\right \}_{s,s}(\x,\x)d\x .\nonumber
\end{align} 

Now consider the r.h.s. of (\ref{widomstredar}). Since the magnetic
field multiplying the spin will not change, our notation will only
refer to $B_2$. Due to the stability of the spectrum against small variations of the magnetic field, for sufficiently small 
$\Delta B$, $\Pi^{B_2+\Delta B}$ still exists and
\begin{equation}
\Pi^{B_2+\Delta B}-\Pi^{B_2}=
\frac{i}{2\pi}\int_{\Gamma} ((H_\infty(B_2+\Delta B)-z)^{-1}-
(H_\infty(B_2)-z)^{-1})dz.
\end{equation}
By using the magnetic perturbation theory \cite{N2} with respect to $\Delta B$
(see also the discussion around \eqref{identit3}) one obtains:
\begin{align}
&[\Pi^{B_2+\Delta B}-\Pi^{B_2}](\x,\x)=
\frac{ie \Delta B}{2\pi mc}
\int_{\Gamma} \left\{ \int_{\R^{3}}d\y(H_\infty(B_2)-z)^{-1}
(\x,\y)\right . \\
& \left . \frac{}{} [{\bf P}_\y(B_2)\cdot {\bf A}(\y-\x)]
(H_\infty(B_2)-z)^{-1}(\y,\x)\right \}dz +\mathcal{O}((\Delta B)^{2}).\nonumber
\end{align}
Now a very important identity is (see (\ref{ident1}), \eqref{nocontr100} and \eqref{nocontr200})
\begin{align}\label{comutxcur}
&{\bf A}(\y-\x)
(H_\infty(B_2)-z)^{-1}(\y,\x)=\\
&
-\frac{i}{2m}
{\bf n}_3\wedge
 [(H_\infty(B_2)-z)^{-1}
{\bf P}(B_2)
(H_\infty(B_2)-z)^{-1}](\y,\x).\nonumber
\end{align}
The remainder in $(\Delta B)^2$ will have a smooth integral kernel
after the integration with respect to $z$, hence we obtain:
\begin{align}\label{widomstredar3}
&\frac{\partial}{\partial B_{2}}
\frac{1}{ |\Omega|} 
\int_\Omega \Pi^{B}(\x,\x)d\x=-\frac{e}{4\pi m^{2}c}\\ 
&
\cdot \int_\Omega\left\{\int_{\Gamma}\left [\frac{}{} 
(H_\infty(B)-z)^{-1}P_{1}(B)(H_\infty(B)-z)^{-1}P_{2}(B)
(H_\infty(B)-z)^{-1} \right .\right .\nonumber \\
&\left .\left . \frac{}{}-(H_\infty(B)-z)^{-1}P_{2}(B)
(H_\infty(B)-z)^{-1}P_{1}(B)(H_\infty(B)-z)^{-1})dz\right ]\right \} 
(\x,\x).\nonumber
\end{align}
From (\ref{widomstredar3}) and (\ref{widomstredar20}) we see that 
(\ref{widomstredar}) follows if we can prove that 
one can circularly permute the operators under the integral sign in 
(\ref{widomstredar3}). One can prove this by interpreting 
(\ref{widomstredar3}) as the thermodynamic limit of the 
corresponding expression on finite volume and then using the
invariance of the trace under cyclic  permutations. 
Alternatively one can prove it directly and in what 
follows we outline the proof. 
 
Due to the smoothing effect of the integral with respect to $z$, we
can always restrict ourselves to considering a product of only 
two integral operators which commute
with the discrete magnetic translations, and have kernels
$e^{ib \phi_{{\bf a}}(\x,\y)}K_1(\x,\y)$ and $e^{ib \phi_{{\bf
      a}}(\y,\x')}K_2(\y,\x')$. We therefore look at an absolutely convergent 
integral of the form (the anti-symmetric magnetic phases
disappear when we look at the diagonal, see \eqref{intkerb2}):
\begin{equation}
\int_\Omega d\x\int_{\R^{3}}d\y K_1 (\x,\y)K_2(\y,\x)
\end{equation}
with
$$
K_{1,2} (\x,\y)=
K_{1,2}(\x+\ggamma,\y+\ggamma).
$$
Then
\begin{align}\label{ciclictrasa}
&\int_{\Omega} d\x\int_{\R^{3}}d\y  K_1 (\x,\y)K_2(\y,\x)=
\sum_{\ggamma \in \mathcal{L}}\int_\Omega
d\x\int_{\Omega}d\y K_1 (\x,\y+\ggamma)K_2(\y+\ggamma,\x) \nonumber \\
&=
\sum_{\ggamma \in \mathcal{L}}\int_\Omega
d\x\int_{\Omega}d\y K_1 (\x-\ggamma,\y)K_2(\y,\x-\ggamma) \nonumber \\ 
& =\int_{\Omega} d\y \int_{\R^{3}}d\x \int_{\R^{3}}K_2(\y,\x)K_1 (\x,\y)
\end{align}
which gives the needed ``trace cyclicity'' and the theorem is
proved.\qed 
\vspace{0.5cm}

We now turn to the question whether the limit in (\ref{widomstredar})
actually vanishes as is suggested by some heuristic arguments (see e.g. \cite{R2}). We start by recalling some results about Wannier functions. Let $\sigma_{0}(B_{0})$ be an isolated part of the spectrum of $H_{\infty}(B_{0})$ and $\Pi^{B_{0}}_{0}$
the corresponding spectral projection. We say that  $\Pi^{B_{0}}_{0}$ has a basis of exponentially localized (magnetic) Wannier functions if there exist
$\alpha >0$, $w_{j}\in L^2(\R^3)\oplus L^2(\R^3),\; j=1,2,...,p<\infty$ satisfying
(we denote by $w_j(\x,s)\in\mathbb{C}$ 
the values of $w_j$ in $\x\in\R^3$ and $s\in\{1,2\}$)
\begin{equation}\label{explocwan}
\sum_{s=1}^{2}\int_{\R^{3}}|w_{j}(\x,s)|^{2}e^{2\alpha |\x|}d\x\leq M<\infty,
\end{equation}
such that the set of functions 
$\left \{ w_{j,\ggamma} \right \}_{j=1,2,...p,\;\ggamma \in \mathcal{L}}$
with
$$ w_{j,\ggamma}(\x,s)= e^{ib \phi_{{\bf a}}(\x,\ggamma)} w_{j}(\x-\ggamma, s)$$ is a basis in the range
of the projection 
 $\Pi^{B_{0}}_{0}(L^{2}(\R^{3})\oplus L^2(\R^3))$. If the spin is   neglected, it has been proved in \cite{N1} that the existence of bases of exponentially localized Wannier functions is stable against small values of the magnetic field (i.e. $B_{0}=0$). More precisely, if $\sigma_{0}$ is an isolated part of the spectrum of $-\Delta +V$ and the corresponding subspace has a basis of  exponentially localized Wannier functions then, for sufficiently small $B$,  $\sigma_{0}(B)$
is still isolated and the corresponding spectral subspace has a basis  of exponentially localized magnetic Wannier functions. The methods in \cite{NN}
together with the magnetic perturbation theory \cite{CN}, \cite{Cor},
\cite{N2} 
allow one to generalize the above result to arbitrary $B_{0}$ and presence of the
 spin (as far as the spin-orbit term  is sufficiently small)
 \cite{CNN}. 
Now the existence of  exponentially localized magnetic Wannier
functions for an isolated part of the spectrum and for 
the value of the magnetic field $B_2$ in an interval around $B_0$
allows one to write:
\begin{align}
& \sum_{s=1}^{2}\int_{\Omega}\Pi^{B}_{0;s,s}(\x,\x)d\x = \int_{\Omega}\sum_{j,\ggamma}
\sum_{s=1}^{2}|w_{j,\ggamma}(\x,s)|^{2}d\x \nonumber \\
&=
 \sum_{j,\ggamma}\sum_{s=1}^{2} \int_{\Omega}d\x
|w_{j}(\x-\ggamma,s)|^{2}=
\sum_{j=1}^p\sum_{s=1}^{2} \int_{\R^{3}}d\x
|w_{j}(\x,s)|^{2}= p.
\end{align}
Thus the integrated density of states corresponding to the Fermi
projection is constant in $B_2$ in a small interval around $B_0$, 
hence this band gives no contribution in the r.h.s. of
(\ref{widomstredar}). 

For small fields, the above discussion can be summarized in:
\begin{theorem}\label{wanierx}
Suppose $(d_1, d_2)\subset \rho (-\Delta +V)$, $d_2>d_1$, and that the
spectral subspace corresponding to $(-\infty, d_1]$ admits a basis of 
exponentially localized Wannier functions. 
Suppose that the spin-orbit interaction (see \eqref{impuls}) 
is small enough such that as $c^{2}$ decreases from $\infty$ to its
actual value, we have that $\frac{d_1 +d_2}{2}\in
\rho(H_{\infty}(0))$. Then for sufficiently small $B$:
\begin{equation}\label{widomstredarf}
\lim_{\omega \rightarrow 0}\lim_{T \rightarrow 0} \sigma_{12}(B,T,\omega)=0.
\end{equation}
\end{theorem}

\vspace{0.5cm}

In particular all the derivatives of $\sigma_{12}(B,0)$ vanish for $B=0$, and this substantiates  Roth's result (formula (50) in \cite{R2})
for the first order correction in $B$ at zero frequency.

\section{ A closed formula for free electrons} \label{sectiuneacinci}

If $V=0$ it turns out that the conductivity tensor can be
explicitly computed for all values of $B$ and $\omega$. The formula
does not depend on whether we work in two or three dimensions. More
precisely, we will show in this section that
\begin{equation}\label{freeatlast}
\sigma_{12} (B,\omega)=\frac{e^3 n}{m^2 c }
\frac{B}{\omega^2-\frac{B^2 e^2}{m^2 c^2}}, 
\end{equation}
where $n=n(T,\mu,B)$ is the grand-canonical density. The formula
(\ref{freeatlast}) is well known in classical physics and goes back at
least to Drude but we are not aware of a known fully quantum derivation. The
coincidence of classical and quantum formulas can be understood taking
into account that the Hamiltonians involved (choose the symmetric
gauge) are quadratic and it is known that for this class of operators 
classical and
quantum computations coincide in many instances. While it is possible
to derive (\ref{freeatlast})
by using the explicit form  of the Green function or alternatively
of eigenvalues and eigenprojections for the Landau Hamiltonian (see e.g. 
\cite{JP}) we shall obtain it below only using resolvent
and commutation identities.

Let us only notice that when $\omega=0$ we re-obtain formula (18) in
\cite{Str}, while for a fixed frequency we get
$$\frac{\partial\sigma_{12}}{\partial B} (0,\omega)=\frac{e^3 n}{m^2 c \omega^2}$$
which is ``the high frequency limit'' or what Roth also calls ``the free 
electron Faraday effect'' in formula (51) from \cite{R2}.

We begin by listing a few identities which are valid for a free electron on
the entire space.
\begin{align}\label{free2}
 & i[P_1(B),P_2(B)] = \frac{B\;e}{c} \nonumber \\
 &
 i[H_\infty(B),P_1(B)]=-\frac{B\;e}{m\;c}\;P_2(B),\nonumber
 \\
 & i[H_\infty(B),P_2(B)]=\frac{B\;e}{m\; c}\;P_1(B), \nonumber \\
 & [H_\infty(B),[H_\infty(B),P_1(B)]] =
\frac{B^2 e^2}{m^2 c^2}P_1(B), \\
 & [H_\infty(B),[H_\infty(B),P_2(B)]] =
\frac{B^2 e^2}{m^2 c^2}P_2(B).\nonumber
\end{align}
Next, since in this case $\mathcal{A}_{s,s}^\infty(\x,\x)$
does not depend upon $\x$ one has
\begin{align}\label{free1}
&  a(B,\omega) =-\frac{1}{2\pi} \sum_{s=1}^2 \\
&\left \{ \int_{\Gamma_\omega} dz 
{f}_{FD}(z) 
\left [ P_1(B)(H_\infty(B)-z)^{-1}P_2(B) 
(H_\infty(B)-z-\omega)^{-1}\right .\right .\nonumber \\
&+ \left .\frac{}{}\left . z \rightarrow z-\omega \right ]\right \}(\vec{0},s;\vec{0},s).\nonumber 
\end{align}
Commuting $(H_\infty(B)-z-\omega)^{-1}$ with $P_2(B)$ in the first
term, and $P_1(B)$ with $(H_\infty(B)-z+\omega)^{-1}$ in the second
one, we obtain
\begin{align}\label{free3}
& a(B,\omega) =-\frac{1}{2\pi |\Omega|} 
 \sum_{s=1}^2 \nonumber \\
 & \left \{ \int_{\Gamma_\omega} dz 
{f}_{FD}(z) 
\left [ P_1(B)(H_\infty(B)-z)^{-1}
(H_\infty(B)-z-\omega)^{-1}P_2(B) \frac{}{}\right .\right .\nonumber
\\
&+ P_1(B)(H_\infty(B)-z)^{-1}
(H_\infty(B)-z-\omega)^{-1}\nonumber \\
&\cdot [H_\infty(B),P_2(B)](H_\infty(B)-z-\omega)^{-1}\nonumber
\\
&+ (H_\infty(B)-z+\omega)^{-1}P_1(B)
P_2(B) 
(H_\infty(B)-z)^{-1}\nonumber \\
&+(H_\infty(B)-z+\omega)^{-1}[H_\infty(B),P_1(B)]\nonumber \\ 
&\left . \left .\frac{}{}\cdot (H_\infty(B)-z+\omega)^{-1}
P_2(B) 
(H_\infty(B)-z)^{-1} \right ] \right
\}(\vec{0},s;\vec{0},s)\nonumber \\
&= I+II+III+IV.
\end{align}
Now $I+III$ can easily be computed. Indeed, by cyclic permutations
(see \eqref{ciclictrasa}) one
can cluster the two resolvents and then by the resolvent identity
\begin{equation}\label{identr}
(A-z_1)^{-1}(A-z_2)^{-1}=(z_1-z_2)^{-1}[(A-z_1)^{-1}-(A-z_2)^{-1}],
\end{equation}
one obtains four terms. Two of them vanish after the integration over
$z$ due to the analyticity of the integrand while the other two give
\begin{align}\label{free4}
& I+III=\frac{1}{2\pi }  \sum_{s=1}^2 \\
& \left \{ \int_{\Gamma_\omega} dz 
{f}_{FD}(z) 
  \frac{1}{\omega}[P_2(B),P_1(B)](H_\infty(B)-z)^{-1} 
\right \}(\vec{0},s;\vec{0},s)\nonumber\\
&= \frac{B\;e}{\omega }\sum_{s=1}^2
\{f_{FD}(H_\infty(B))\}(\vec{0},s;\vec{0},s) =:
\frac{B\;e}{\omega }n(T,\mu,B).
\end{align}
In an analogous manner
\begin{align}\label{free6}
& III+IV=\frac{1}{2\pi \omega} \sum_{s=1}^2 \left \{ \int_{\Gamma_\omega} dz 
{f}_{FD}(z) \right .
\\
&\cdot \left \{ 
(H_\infty(B)-z)^{-1}[H_\infty(B),P_2(B)](H_\infty(B)-z-\omega)^{-1}P_1(B)
\right .\nonumber
\\
&-\left .(H_\infty(B)-z)^{-1}[H_\infty(B),P_1(B)](H_\infty(B)-z+\omega)^{-1}
P_2(B) \right \} \nonumber \\
& \left . \frac{}{}\right \}(\vec{0},s;\vec{0},s).\nonumber  
\end{align}
At this point we commute $[H_\infty(B),P_2(B)]$ with
$(H_\infty(B)-z-\omega)^{-1}$ in the first term, and $[H_\infty(B),P_1(B)]$ with
$(H_\infty(B)-z+\omega)^{-1}$ in the second term and use
(\ref{ident1}) again. Some of the terms vanish after performing the
integration over $z$ and the remaining ones write as:
\begin{align}\label{adf2}
&
-\frac{1}{\omega}(H_\infty(B)-z)^{-1}[H_\infty(B),P_2(B)]P_1(B)
\\
&-  \frac{1}{\omega}(H_\infty(B)-z)^{-1}[H_\infty(B),[H_\infty(B),P_2(B)]]
(H_\infty(B)-z-\omega)^{-1}P_1(B)\nonumber \\
&-\frac{1}{\omega}(H_\infty(B)-z)^{-1}[H_\infty(B),P_1(B)]P_2(B) 
\nonumber \\
&- \frac{1}{\omega}(H_\infty(B)-z)^{-1} [H_\infty(B),[H_\infty(B),P_1(B)]]
(H_\infty(B)-z+\omega)^{-1}P_2(B).\nonumber 
\end{align}
Taking into account (\ref{free2}) the first and the third terms in
(\ref{adf2}) combine to
$$-\frac{1}{\omega}(H_\infty(B)-z)^{-1}[H_\infty(B),P_1(B)P_2(B)]$$
which after integration over $z$ is proportional to
\begin{align}\label{singularmoth}
& f_{FD}(H_\infty(B))i[H_\infty(B),P_1(B)P_2(B)]\\
&=\frac{B\;e}{m\;
  c}\;\left \{f_{FD}(H_\infty(B))P_1(B)^2-f_{FD}(H_\infty(B))P_2(B)^2\right \}, \nonumber
\end{align}
where we used the second and third identities in
\eqref{free2}. Consider the unitary operator $U$ which implements the
coordinate change $(Uf)(x_1,x_2,x_3)= f(x_2,-x_1,x_3)$. Then one can
prove that $UP_1(B)U^*=-P_2(B)$, $UP_2(B)U^*=P_1(B)$ and 
$UH_\infty(B)U^*=H_\infty(B)$. This implies that 
$$Uf_{FD}(H_\infty(B))P_1(B)^2U^*=f_{FD}(H_\infty(B))P_2(B)^2.$$
Since both operators have a smooth integral kernel, and because the
rotation with $U$ does not change the diagonal value of the integral
kernel on the left hand side, 
it means that the contribution given by \eqref{singularmoth}
is zero.

Therefore we only remain with the second and fourth terms in
(\ref{adf2}).
 Using (\ref{free2}), they become:
\begin{align}\label{adf3}
&-  \frac{B^2 e^2}{m^2 c^2 \omega}(H_\infty(B)-z)^{-1}P_2(B)
(H_\infty(B)-z-\omega)^{-1}P_1(B)\nonumber \\
&- \frac{B^2 e^2}{m^2 c^2 \omega}(H_\infty(B)-z)^{-1}P_1(B)
(H_\infty(B)-z+\omega)^{-1}P_2(B).
\end{align}
Using once more the cyclicity of the trace and comparing with the
starting point (\ref{free1}), we obtain the
remarkable identity
\begin{equation}\label{adf4}
II+IV= \frac{B^2 e^2}{m^2 c^2 \omega^2} a(B,\omega).
\end{equation}
Putting together (\ref{free3}), (\ref{free4}), and
(\ref{adf4}), we obtain the equation:
$$a(B,\omega)=\frac{B\;e}{c \omega}n +
\frac{B^2 e^2}{m^2 c^2 \omega^2} a(B,\omega),$$
which gives (\ref{freeatlast}) (see (\ref{condupevol})).

\section{Magnetic perturbation theory and the linear term in $B$}\label{sectiuneasapte}

When $V\not=0$ it is no longer possible to obtain a closed formula for
$\sigma_{12}(B,\omega)$. Since in most physical applications the
external magnetic field can be considered weak, 
an expansion in $B$ up to the first or second
order would be sufficient. In this section we show that
$a_L(B,\omega )$ has an expansion in $B$ to any order and write down
the expressions of the first two terms. The first one gives the
transverse conductivity at zero magnetic field and the 
second which is linear in $B$ provides the Verdet constant.
From (\ref{calA}) and (\ref{ik2}) (in what follows by $tr$ we mean the trace over the
spin variable):
\begin{align}\label{identit2}
& a_L(B,\omega )= -\frac{1}{2\pi |\Lambda_L|}\int_{\Lambda_L} d\x 
\left \{{\rm tr}\int_{\Gamma_\omega}dz  f_{FD}(z)\right. \\
&\cdot \int_{\Lambda_L}d{\bf u} 
\left \{\left [(P_{\x,1} (0)-b 
A_1(\x-{\bf u}))K_{L}(\x,{\bf u}; z )
\right ]\right .
\nonumber \\ 
&\cdot  
\left [(P_{{\bf u},2} (0)-b 
A_2({\bf u}-\x'))K_{L}({\bf u},\x'; z+\omega)\right ]
\nonumber\\
&+ \left .\left [(P_{\x,1} (0)-b
A_1(\x-{\bf u}))K_{L}(\x,{\bf u}; z-\omega)
\right ]\right .
\nonumber \\ 
&\cdot  \left . \left .\frac{}{} \left .
\left [(P_{{\bf u},2} (0)-b  
A_2({\bf u}-\x'))K_{L}({\bf u},\x'; z)\right ]
\right \}\right \}\right \vert_{\x=\x'}\nonumber
\end{align}

Let us mention here that 
one cannot interchange the order of the above integrals. First one
performs the integral with respect to ${\bf u}$, 
then the integral in $z$, then we can put $\x=\x'$ since the resulting kernel 
is smooth, and finally one integrates with respect to $\x$ over $\Lambda_L$.  

When considering the expansion   in $b$ of  $a_L(B,\omega )$ we are
left with the problem of the expansion of
$K_{\Lambda_L}(\x,\y;\zeta)$ . This expansion is provided by the 
magnetic perturbation theory as developed in \cite{N2}. 
Following the steps in  \cite{N2} in the case at hand one obtains: 
                 
\begin{align}\label{identit3}
& K_{L}(\x,\y; z)=G_{L}^{(0)}(\x,\y; z)\\
&+
\frac{b}{m}\int_{\Lambda_L}G_{L}^{(0)}(\x,{\bf u}; z)
\left [{\bf P}_{{\bf u}}(0)\cdot \A({\bf u}-\y)
G_{L}^{(0)}({\bf u},\y; z)\right ]d{\bf u}\nonumber \\
&+ b 
\frac{ g c\mu_b}{e}\int_{\Lambda_L}G_{L}^{(0)}(\x,{\bf u}; z)
\sigma_3
G_{L}^{(0)}({\bf u},\y; z)d{\bf u} +{\cal O}(b^2)
\nonumber\\
&= G_{L}^{(0)}(\x,\y; z)+ b 
G_{L}^{(orbit)}(\x,\y; z)+ b 
G_{L}^{(spin)}(\x,\y; z)+{\cal O}(b^2).\nonumber
\end{align}
The above integrands are matrices in the spin variable, that is why
the spin does 
not appear explicitly. The error term ${\cal O}(b^2)$ can also be fully controlled with the 
magnetic perturbation theory (actually arbitrary order terms can be computed; see \cite{N1} for details).
Plugging the expansion (\ref{identit3}) into (\ref{identit3})
 and collecting the terms of zero and first order one obtains
\begin{align}\label{identit4}
a_L(B,\omega )=a_L(0,\omega)+ba_{L,1}(\omega)
+
\mathcal{O}(b^2),
\end{align}
where the zeroth order term is:
\begin{align}\label{alo}
& a_{L}(0,\omega) =
-\frac{1}{2\pi |\Lambda_L|}\int_{
\Lambda_L} d\x \left\{
{\rm tr}\int_{\Gamma_\omega}dz  f_{FD}(z)\right . \\
&\left . \left . \frac{}{}\cdot  \{ P_1(0)
(H_{L}(0)- z)^{-1}P_{2}(0)(H_{L}(0)- z+\omega)^{-1}+ (z\to z-\omega) \}
 \right \} \right \vert_{\x=\x'},\nonumber
\end{align}
while the first order correction reads as:
\begin{equation}\label{identit50}
a_{L,1}(\omega)=a_{L,1}^{orbit}(\omega)+a_{L,1}^{spin}(\omega),
\end{equation}
where 
\begin{align}\label{identit5}
& a_{L,1}^{orbit}(\omega)=-\frac{1}{2\pi |\Lambda_L|}\int_{
\Lambda_L} d\x 
\left \{{\rm tr}\int_{\Gamma_\omega}dz  f_{FD}(z)\right. \\
&\cdot \int_{\Lambda_L}d{\bf u} 
\left \{-\left [
A_1(\x-{\bf u})G_{L}^{(0)}(\x,{\bf u}; z )
\right ]
\left [P_{{\bf u},2} (0)G_{L}^{(0)}
({\bf u},\x'; z+\omega)\right ]\right .
\nonumber\\
&-
\left [P_{{\bf x},1} (0)G_{L}^{(0)}
(\x,{\bf u}; z)\right ]\left [ 
A_2({\bf u}-\x')G_{L}^{(0)}({\bf u},\x'; z+\omega)\right ]
\nonumber \\
&+\left [P_{{\bf x},1} (0)G_{L}^{(orbit)}
(\x,{\bf u}; z)\right ]\left [
P_{{\bf u},2} (0)G_{L}^{(0)}
({\bf u},\x'; z+\omega)\right ]\nonumber \\
&+
\left [P_{{\bf x},1} (0)G_{L}^{(0)}
(\x,{\bf u}; z)\right ]\left [
P_{{\bf u},2} (0)G_{L}^{(orbit)}
({\bf u},\x'; z+\omega)\right ]\nonumber \\
&+ \left.\left .\left . \frac{}{}(z\to z-\omega) \right \} \right \}\right 
\vert_{\x=\x'},\nonumber
\end{align}
\begin{align}\label{identit6}
& a_{L,1}^{spin}(\omega)=-\frac{1}{2\pi |\Lambda_L|}\int_{\Lambda_L} d\x 
\left \{{\rm tr}\int_{\Gamma_\omega}dz  f_{FD}(z)\right. \\
&\cdot \int_{\Lambda_L}d{\bf u} 
\left \{
\left [P_{{\bf x},1} (0)G_{L}^{(spin)}
(\x,{\bf u}; z)\right ]\left [
P_{{\bf u},2} (0)G_{L}^{(0)}
({\bf u},\x'; z+\omega)\right ]\right .\nonumber \\
&+
\left [P_{{\bf x},1} (0)G_{L}^{(0)}
(\x,{\bf u}; z)\right ]\left [
P_{{\bf u},2} (0)G_{L}^{(spin)}
({\bf u},\x'; z+\omega)\right ]\nonumber \\
&+\left .\left .\left . \frac{}{}(z\to z-\omega) \right \} \right \} \right
\vert_{\x=\x'}.\nonumber
\end{align}
Now consider the expression ${\bf A}(\x-\y)G_{L}^{(0)}(\x,\y;
z)$ appearing in the formula for 
$a_{L,1}(\omega)$. Observing that it represents  a commutator (see (\ref{ident1})) one has the identity
\begin{align}\label{identit7}
&  {\bf A}(\x-\y)G_{L}^{(0)}(\x,\y; z)=\left ( \frac{1}{2} {\bf n}_3
\wedge (\x-\y)\right )G_{L}^{(0)}(\x,\y; z)\nonumber \\
&=\left (\frac{1}{2} {\bf n}_3
\wedge \left [{\bf X},(H_{L}(0)- z)^{-1}\right ]\right )(\x,\y)\nonumber \\ 
&=
-\frac{i}{2m}\{(H_{L}(0)- z)^{-1}({\bf n}_3\wedge P )
(H_{L}(0)- z)^{-1}\}(\x,\y),
\end{align}
where ${\bf X}$ denotes the multiplication operator with $\x$. By a
straightforward (but somewhat tedious) 
computation one arrives at: 
\begin{equation}\label{agama1}
a_{L,1}(\omega)=a_{L,1}^{orbit,1}(\omega)+a_{L,1}^{orbit,2}(\omega)+
a_{L,1}^{spin}(\omega)
\end{equation}
where
\begin{align}\label{identit14}
a_{L,1}^{orbit,1}(\omega)&=\frac{i}{4m\pi\omega |\Lambda_L|}\int_{\Lambda_L} d\x 
\left \{{\rm tr}\int_{\Gamma_\omega}dz  f_{FD}(z)\right. \\
&\cdot\left [ \frac{}{}\sum_{\alpha=1}^2P_\alpha(0)
(H_{L}(0)- z)^{-1}P_{\alpha}(0)(H_{L}(0)- z-\omega)^{-1}\right .
\nonumber\\
&+ \sum_{\alpha=1}^2P_\alpha(0)
(H_{L}(0)- z)^{-1}P_{\alpha}(0)(H_{L}(0)- z+\omega)^{-1}
\nonumber\\
&\left . \left .\frac{}{}- \sum_{\alpha=1}^2
P_{\alpha}(0)
(H_{L}(0)- z)^{-1}P_\alpha(0)
(H_{L}(0)- z)^{-1}\right ]
\right \} (\x,\x)\nonumber,
\end{align}
\begin{align}\label{identit17}
 a_{L,1}^{orbit,2}(\omega)
&=\frac{i}{4\pi m^2|\Lambda_L|}\int_{\Lambda_L} d\x 
\left \{{\rm tr}\int_{\Gamma_\omega}dz  f_{FD}(z)\right. \\
&\cdot \left \{
-P_{1}(0)(H_{L}(0)- z)^{-1}P_{1}(0)
(H_{L}(0)- z)^{-1}\right . \nonumber \\
&\cdot P_{2}(0)(H_{L}(0)-
z)^{-1}
P_{2}(0)(H_{L}(0)- z-\omega)^{-1}\nonumber \\
&+ P_{1}(0)(H_{L}(0)- z)^{-1}P_{2}(0)
(H_{L}(0)- z)^{-1}\nonumber \\
&\cdot P_{1}(0)(H_{L}(0)-z)^{-1}
P_{2}(0)(H_{L}(0)- z-\omega)^{-1}\nonumber \\
&-
P_{1}(0)(H_{L}(0)- z+\omega)^{-1}P_{1}(0)
(H_{L}(0)- z+\omega)^{-1}\nonumber \\ 
&\cdot P_{2}(0)(H_{L}(0)-
z+\omega)^{-1}
P_{2}(0)(H_{L}(0)- z)^{-1}\nonumber \\
&+ P_{1}(0)(H_{L}(0)- z+\omega)^{-1}P_{2}(0)
(H_{L}(0)- z+\omega)^{-1}\nonumber \\
&\cdot P_{1}(0)(H_{L}(0)-
z+\omega)^{-1}
P_{2}(0)(H_{L}(0)- z)^{-1}\nonumber \\
&- P_{1}(0)(H_{L}(0)- z)^{-1}P_{2}(0)
(H_{L}(0)- z-\omega)^{-1}\nonumber \\
&\cdot P_{1}(0)(H_{L}(0)-
z-\omega)^{-1}
P_{2}(0)(H_{L}(0)- z-\omega)^{-1}\nonumber \\
&+ P_{1}(0)(H_{L}(0)- z)^{-1}P_{2}(0)
(H_{L}(0)- z-\omega)^{-1}\nonumber \\
&\cdot P_{2}(0)(H_{L}(0)-
z-\omega)^{-1}
P_{1}(0)(H_{L}(0)- z-\omega)^{-1}\nonumber \\
&- P_{1}(0)(H_{L}(0)- z+\omega)^{-1}P_{2}(0)
(H_{L}(0)- z)^{-1}\nonumber \\
&\cdot P_{1}(0)(H_{L}(0)-
z)^{-1}
P_{2}(0)(H_{L}(0)- z)^{-1}\nonumber \\
&+ P_{1}(0)(H_{L}(0)- z+\omega)^{-1}P_{2}(0)
(H_{L}(0)- z)^{-1}\nonumber \\
& \left .\frac{}{}\left . \cdot P_{2}(0)(H_{L}(0)-
z)^{-1}
P_{1}(0)(H_{L}(0)- z)^{-1}\right \} \right
\}(\x,\x),\nonumber
\end{align}
and
\begin{align}\label{identitnoua1}
& a_{L,1}^{spin}(\omega)=-\frac{g c\mu_b}{2e\pi |\Lambda_L|}\int_{\Lambda_L} d\x 
\left \{{\rm tr}\int_{\Gamma_\omega}dz  f_{FD}(z)\right. \\
&\cdot  
\left \{
\left [P_{1}(0)(H_{L}(0)- z)^{-1}\sigma_3(H_{L}(0)- z)^{-1}
P_{2}(0)(H_{L}(0)- z-\omega)^{-1}\right ]\right .\nonumber \\
&+
\left [P_{1}(0)(H_{L}(0)- z)^{-1}
P_{2}(0)(H_{L}(0)-
z-\omega)^{-1}\sigma_3(H_{L}(0)- z-\omega )^{-1}\right ]\nonumber
\\
&+\left [P_{1}(0)(H_{L}(0)-
  z+\omega)^{-1}\sigma_3(H_{L}(0)- z+\omega )^{-1}
P_{2}(0)(H_{L}(0)- z)^{-1}\right ]\nonumber
\\
&\left . \frac{}{}+
\left [P_{1}(0)(H_{L}(0)- z+\omega)^{-1}
P_{2}(0)(H_{L}(0)-
z)^{-1}\sigma_3(H_{L}(0)- z )^{-1}\right ]\}\right \} (\x,\x).\nonumber
\end{align}

 \section{The periodic case}\label{sectiuneaopt}

Now consider the case when $V$ is periodic. In this case, after taking
the thermodynamic limit one can replace 
(see (\ref{stone8})) $\frac{1}{ |\Lambda_L|}\int_{\Lambda_L}$ with
 $\frac{1}{ |\Omega|}\int_{\Omega}$ and rewrite
(\ref{identit14})-(\ref{identitnoua1}) as integrals over the Brillouin zone
\begin{align}\label{identit141}
a_{\infty, 1}^{orbit,1}(\omega)&=\frac{i}{4m\pi\omega |\Omega|}
\int_{\Omega*}d\bk\int_{\Omega} d\x 
\left \{{\rm tr}\int_{\Gamma_\omega}dz  f_{FD}(z)\right. \\
&\cdot \sum_{\alpha=1}^2 (p_{\alpha}+k_{\alpha})
(h(\bk)- z)^{-1}(p_{\alpha}+k_{\alpha})(h(\bk)- z-\omega)^{-1}
\nonumber\\
&+ \sum_{\alpha=1}^2 (p_{\alpha}+k_{\alpha})
(h(\bk)- z)^{-1}(p_{\alpha}+k_{\alpha})(h(\bk)- z+\omega)^{-1}
\nonumber\\
&- \sum_{\alpha=1}^2
(p_{\alpha}+k_{\alpha})
(h(\bk)- z)^{-1}P_\alpha(0)
(h(\bk)- z)^{-1}
\left . \frac{}{} \right \} (\x,\x)\nonumber,
\end{align}

\begin{align}\label{identit171}
a_{\infty,1}^{orbit,2}(\omega)
&=\frac{i}{4\pi m^2 |\Omega|}\int_{\Omega*}d\bk\int_{\Omega} d\x 
\left \{{\rm tr}\int_{\Gamma_\omega}dz  f_{FD}(z)\right. \\
&\cdot \{
-(p_1 +k_1)(h(\bk)- z)^{-1}(p_1 +k_1)
(h(\bk)- z)^{-1}\nonumber \\
&\cdot (p_2 +k_2)(h(\bk)- z)^{-1}
(p_2 +k_2)(h(\bk)- z-\omega)^{-1} \nonumber \\
&+ (p_1 +k_1)(h(\bk)- z)^{-1}(p_2 +k_2)
(h(\bk)- z)^{-1}\nonumber \\ 
&\cdot (p_1 +k_1)(h(\bk)-
z)^{-1}
(p_2 +k_2)(h(\bk)- z-\omega)^{-1}\nonumber \\
&-
(p_1 +k_1)(h(\bk)- z+\omega)^{-1}(p_1 +k_1)
(h(\bk)- z+\omega)^{-1}\nonumber \\
&\cdot (p_2 +k_2)(h(\bk)-z+\omega)^{-1}
(p_2 +k_2)(h(\bk)- z)^{-1}\nonumber \\
&+ (p_1 +k_1)(h(\bk)- z+\omega)^{-1}(p_2 +k_2)
(h(\bk)- z+\omega)^{-1}\nonumber \\
&\cdot (p_1 +k_1)(h(\bk)-
z+\omega)^{-1}
(p_2 +k_2)(h(\bk)- z)^{-1}\nonumber \\
&- (p_1 +k_1)(h(\bk)- z)^{-1}(p_2 +k_2)
(h(\bk)- z-\omega)^{-1}\nonumber \\
&\cdot (p_1 +k_1)(h(\bk)-
z-\omega)^{-1}
(p_2 +k_2)(h(\bk)- z-\omega)^{-1}\nonumber \\
&+ (p_1 +k_1)(h(\bk)- z)^{-1}(p_2 +k_2)
(h(\bk)- z-\omega)^{-1}\nonumber \\
&\cdot (p_2 +k_2)(h(\bk)-
z-\omega)^{-1}
(p_1 +k_1)(h(\bk)- z-\omega)^{-1}\nonumber \\
&- (p_1 +k_1)(h(\bk)- z+\omega)^{-1}(p_2 +k_2)
(h(\bk)- z)^{-1}\nonumber \\
&\cdot (p_1 +k_1)(h(\bk)-
z)^{-1}
(p_2 +k_2)(h(\bk)- z)^{-1}\nonumber \\
&+ (p_1 +k_1)(h(\bk)- z+\omega)^{-1}(p_2 +k_2)
(h(\bk)- z)^{-1}\nonumber \\
&\cdot (p_2 +k_2)(h(\bk)-
z)^{-1}(p_1 +k_1)(h(\bk)- z)^{-1}\} \left .\frac{}{}\right \}
(\x,\x),\nonumber
\end{align}
and
\begin{align}\label{identitnoua11}
& a_{\infty, 1}^{spin}(\omega)=-\frac{g c\mu_b}{2e\pi |\Omega|}\int_{\Omega*}d\bk\int_{\Omega} d\x 
\left \{{\rm tr}\int_{\Gamma_\omega}dz  f_{FD}(z)\right. \\
&\cdot   
\left \{
\left [(p_1 +k_1)(h(\bk)- z)^{-1}\sigma_3(h(\bk)- z)^{-1}
(p_2 +k_2)(h(\bk)- z-\omega)^{-1}\right ]\right .\nonumber \\
&+
\left [(p_1 +k_1)(h(\bk)- z)^{-1}
(p_2 +k_2)(h(\bk)-
z-\omega)^{-1}\sigma_3(h(\bk)- z-\omega )^{-1}\right ]\nonumber
\\
&+\left [(p_1 +k_1)(h(\bk)-
  z+\omega)^{-1}\sigma_3(h(\bk)- z+\omega )^{-1}
(p_2 +k_2)(h(\bk)- z)^{-1}\right ]\nonumber
\\
&+
\left [(p_1 +k_1)(h(\bk)- z+\omega)^{-1}
(p_2 +k_2)(h(\bk)-
z)^{-1}\sigma_3(h(\bk)- z )^{-1}\right ]\} \left .\frac{}{}\right \}(\x,\x).\nonumber
\end{align}

 Finally, for the convenience of the reader only interested in applying
 the theory to the case when one assumes that the Bloch bands and
 functions are known (as for example from Kohn-Luttinger type models), we write
(\ref{identit141})-(\ref{identitnoua11}) in terms of Bloch functions
and energies. The important thing here is that no derivatives with respect to the quasi-momentum appear.
With  the usual notation (here $\langle,\rangle$ denotes the scalar
product over the spin variables):
\begin{align}\label{kerl3}
\hat{\pi}_{ij}(\alpha,\bk) =\int_{\Omega}\langle u_i(\x,\bk), (p_\alpha+\bk_\alpha) u_j(\x,\bk)\rangle d\x,
\end{align}
and after some rearrangements, the terms coming from the orbital
magnetism are:
\begin{align}\label{kerlfinal2}
a_{\infty,1}^{orbit,1}(\omega)& = \frac{1}{2 m \omega (2\pi)^3}
\sum_{\alpha=1}^2\int_{\Omega^*}d\bk \left\{ \sum_{j \geq 1}
|\hat{\pi}_{jj}(\alpha,\bk)|^2 f_{FD}'(\lambda_j(\bk))\right.
\\  
&-\omega^2\left .\sum_{j\neq l}
|\hat{\pi}_{lj}(\alpha,\bk)|^2
\frac{f_{FD}(\lambda_j(\bk))-f_{FD}(\lambda_l(\bk))}
{[(\lambda_j(\bk)-\lambda_l(\bk))^2-\omega^2](\lambda_j(\bk)-\lambda_l(\bk))}
\right \},\nonumber 
\end{align}
\begin{align}\label{idey1}
 a_{\infty,1}^{orbit,2}(\omega)&=
\frac{1}{2 m^2 (2\pi)^3}\int_{\Omega^*}d\bk 
\sum_{n_1,n_2,n_3,n_4\geq 1}
\frac{1}{2\pi i}\int_{\Gamma_\omega} dz f_{FD}(z)\\
& \left \{
  \frac{\hat{\pi}_{n_4 n_1}(1,\bk)\hat{\pi}_{n_1 n_2}(1,\bk) 
\hat{\pi}_{n_2 n_3}(2,\bk)  
\hat{\pi}_{n_3 n_4}(2,\bk)}{(z-\lambda_{n_1}(\bk))
(z-\lambda_{n_2}(\bk))(z-\lambda_{n_3}(\bk))(z+\omega -\lambda_{n_4}(\bk))}\right . \nonumber\\
&
-\frac{\hat{\pi}_{n_4 n_1}(1,\bk)\hat{\pi}_{n_1 n_2}(2,\bk) 
\hat{\pi}_{n_2 n_3}(1,\bk)  
\hat{\pi}_{n_3 n_4}(2,\bk)}{
(z-\lambda_{n_1}(\bk))
(z-\lambda_{n_2}(\bk))(z-\lambda_{n_3}(\bk))(z+\omega -\lambda_{n_4}(\bk))} \nonumber\\
&
+\frac{\hat{\pi}_{n_4 n_1}(1,\bk)\hat{\pi}_{n_1 n_2}(1,\bk) 
\hat{\pi}_{n_2 n_3}(2,\bk)  
\hat{\pi}_{n_3 n_4}(2,\bk)}{(z-\omega-\lambda_{n_1}(\bk))
(z-\omega-\lambda_{n_2}(\bk))(z-\omega- 
\lambda_{n_3}(\bk))(z -\lambda_{n_4}(\bk))}\nonumber\\
&
-\frac{\hat{\pi}_{n_4 n_1}(1,\bk)\hat{\pi}_{n_1 n_2}(2,\bk) 
\hat{\pi}_{n_2 n_3}(1,\bk)  
\hat{\pi}_{n_3 n_4}(2,\bk)}{(z-\omega-\lambda_{n_1}(\bk))
(z-\omega-\lambda_{n_2}(\bk))(z-\omega- 
\lambda_{n_3}(\bk))(z -\lambda_{n_4}(\bk))}\nonumber\\
&
+\frac{\hat{\pi}_{n_4 n_1}(1,\bk)\hat{\pi}_{n_1 n_2}(2,\bk) 
\hat{\pi}_{n_2 n_3}(1,\bk)  
\hat{\pi}_{n_3 n_4}(2,\bk)}
{(z-\lambda_{n_1}(\bk))
(z+\omega-\lambda_{n_2}(\bk))(z+\omega- 
\lambda_{n_3}(\bk))(z +\omega -\lambda_{n_4}(\bk))}\nonumber \\
& 
-\frac{\hat{\pi}_{n_4 n_1}(1,\bk)\hat{\pi}_{n_1 n_2}(2,\bk) 
\hat{\pi}_{n_2 n_3}(2,\bk)  
\hat{\pi}_{n_3 n_4}(1,\bk)}{(z-\lambda_{n_1}(\bk))
(z+\omega-\lambda_{n_2}(\bk))(z+\omega- 
\lambda_{n_3}(\bk))(z +\omega -\lambda_{n_4}(\bk))}\nonumber \\
& 
+\frac{\hat{\pi}_{n_4 n_1}(1,\bk)\hat{\pi}_{n_1 n_2}(2,\bk) 
\hat{\pi}_{n_2 n_3}(1,\bk)  
\hat{\pi}_{n_3 n_4}(2,\bk) }{(z-\omega-\lambda_{n_1}(\bk))
(z-\lambda_{n_2}(\bk))(z- 
\lambda_{n_3}(\bk))(z  -\lambda_{n_4}(\bk))}\nonumber \\
& \left .
-\frac{\hat{\pi}_{n_4 n_1}(1,\bk)\hat{\pi}_{n_1 n_2}(2,\bk) 
\hat{\pi}_{n_2 n_3}(2,\bk)  
\hat{\pi}_{n_3 n_4}(1,\bk)}{(z-\omega-\lambda_{n_1}(\bk))
(z-\lambda_{n_2}(\bk))(z- 
\lambda_{n_3}(\bk))(z  -\lambda_{n_4}(\bk))}  \right  \}.\nonumber
\end{align}

As for the spin contribution, with the  notation
\begin{equation}\label{spyin}
\hat{s}_{ij}(\bk):=\int_\Omega
\langle u_i(\x,\bk),\sigma_3 u_j (\x,\bk)\rangle d\x, 
\end{equation}
one has:
\begin{align}\label{identitnoua2}
a_{\infty, 1}^{spin}(\omega)&=-\frac{g c\mu_b}{(2\pi)^4
  e}\int_{\Omega^*}d\bk 
\sum_{n_1,n_2,n_3\geq 1}
\frac{1}{2\pi i}\int_{\Gamma_\omega} dz f_{FD}(z)\\
&\left \{
\frac{ \hat{\pi}_{n_1 n_2}(1,\bk)\hat{s}_{n_2
  n_3}(\bk)
\hat{\pi}_{n_3 n_1}(2,\bk)}{(\lambda_{n_2}(\bk)-z)(\lambda_{n_3}(\bk)-z)
(\lambda_{n_1}(\bk)-z-\omega)}\right . \nonumber \\
&+\frac{ \hat{\pi}_{n_1 n_2}(1,\bk)\hat{\pi}_{n_2 n_3}(2,\bk)\hat{s}_{n_3
  n_1}(\bk)}
{(\lambda_{n_2}(\bk)-z)(\lambda_{n_3}(\bk)-z-\omega)
(\lambda_{n_1}(\bk)-z-\omega)} \nonumber \\
&+ \frac{ \hat{\pi}_{n_1 n_2}(1,\bk)\hat{s}_{n_2
  n_3}(\bk)
\hat{\pi}_{n_3 n_1}(2,\bk)}{(\lambda_{n_2}(\bk)-z+\omega)(\lambda_{n_3}(\bk)-z+\omega)
(\lambda_{n_1}(\bk)-z)}\nonumber \\
&\left . + \frac{ \hat{\pi}_{n_1 n_2}(1,\bk)\hat{\pi}_{n_2 n_3}(2,\bk)\hat{s}_{n_3
  n_1}(\bk)}{(\lambda_{n_2}(\bk)-z+\omega)(\lambda_{n_3}(\bk)-z)
(\lambda_{n_1}(\bk)-z)} \right \}.\nonumber 
\end{align}

\section{Conclusions}\label{sectiuneanoua}

We presented in the present paper a method which shed new light on the quantum
dynamics/optical response  in bulk media in the presence of a constant magnetic
field. We applied the gauge invariant magnetic
perturbation theory and gave a clear and very general way of dealing 
with long range magnetic perturbations. 
 
The formal connection with the integer Quantum Hall effect was
established in \eqref{widomstreda20}. Equations  (\ref{agama1})-(\ref{identitnoua1}) and
(\ref{kerl3})-(\ref{identitnoua2}) contain our main
result concerning the Verdet constant and the Faraday effect: 
it gives the linear term in $B$ of the transverse conductivity in
terms of the zero magnetic field Green function. They open the way
of using  the recently developed Green function techniques 
for the calculation of optical 
and magneto-optical properties of solids, to the case when an external
magnetic field is present. 
Our method can be applied to ordered, as well as to random systems (with the
appropriate average over configurations). 
Of course, in the last case one has to assume ergodicity properties in
order to insure convergence of results in the thermodynamic
limit. Layers or other geometries can also be considered.

There are many subtle
and difficult mathematical questions left aside in this paper, as
those related to the thermodynamic limit, the convergence of
infinite series over Bloch bands, and the low frequency limit when the
Fermi energy lies in the spectrum. Another open problem 
is to consider self-interacting
electrons and to investigate the exciton influence on the Faraday
effect. These questions will be addressed elsewhere. 

Our results are not only theoretical. In a future publication we will
use the residue theorem in equations
 (\ref{kerl3})-(\ref{identitnoua2}) to calculate the Verdet
constant for various finite band models, and compare our results with
the existing experimental data. Moreover, our results will be shown to
imply those of Roth \cite{R2} and Nedoluha \cite{Ne}.

\section{Acknowledgments}
This paper has been written at the Department of Mathematical Sciences, Aalborg
University. G.N. kindly thanks the Department for invitation and 
financial support.  

H.C. acknowledges support from the Statens Naturvidenskabelige
Forskningsr{\aa}d grant {\it Topics in Rigorous
  Mathematical Physics}, and partial support through the European Union's IHP
network Analysis $\&$ Quantum HPRN-CT-2002-00277.

\end{document}